\documentclass[twocolumn,9pt]{extarticle} 

\usepackage{authblk}
\expandafter\def\csname ver@subfig.sty\endcsname{}
\usepackage{microtype}

\usepackage[paperheight=285.7mm, paperwidth=215.900mm, left=21.100mm, right=21.100mm, top=15.9mm, bottom = 24mm, footskip=8mm]{geometry}
\usepackage[sort&compress,numbers]{natbib} 
\AtBeginDocument{\def\NAT@space{\relax}}
\usepackage[raggedright,indentafter]{titlesec}
\usepackage{balance}

\titleformat{\section}{\mathversion{bold}\bfseries\large\raggedright}{\thesection}{1em}{}
\titleformat{\subsection}[runin]{\mathversion{bold}\bfseries}{\thesubsection}{1em}{}[.]
\titleformat{\subsubsection}[runin]{\itshape}{\thesubsubsection}{1em}{}[.]

\titlespacing\section{0pt}{14pt plus 3pt minus 2pt}{3pt plus 2pt minus 1pt} 
\titlespacing{\subsection}{\parindent}{12pt plus 3pt minus 2pt}{0.5em}
\titlespacing{\subsubsection}{\parindent}{12pt plus 3pt minus 2pt}{0.5em}

\titleformat{\paragraph}[runin]{\itshape}{\thesubsubsection}{0.5em}{}[.]
\titlespacing{\paragraph}{\parindent}{14pt plus 3pt minus 2pt}{0.5em}

\usepackage[labelfont={sf,bf},hypcap=false]{caption}
\usepackage[hypcap=false,list=true]{subcaption}

\DeclareCaptionTextFormat{boldmath}{\boldmath#1}
\DeclareCaptionTextFormat{sansmath}{#1}

\usepackage{array} 

\usepackage{mathtools}

\usepackage{graphicx} 
\usepackage[dvipsnames,svgnames,x11names]{xcolor}

\usepackage{amsmath,amsthm,amssymb}
\usepackage{enumitem}
\usepackage{siunitx}

\usepackage{nth}
\usepackage{xfrac}

\usepackage{tikz}
\usetikzlibrary{shapes,arrows,chains}
\tikzstyle{block} = [draw, fill=blue!20, rectangle, 
    minimum height=2.5em, minimum width=5em]
\tikzstyle{sum} = [draw, fill=blue!20, circle, node distance=.8cm]
\tikzstyle{input} = [coordinate]
\tikzstyle{output} = [coordinate]
\tikzstyle{pinstyle} = [pin edge={to-,thin,black}]
\usetikzlibrary{positioning}
\usetikzlibrary{patterns}
\usetikzlibrary{calc}
\usetikzlibrary{shapes.misc}
\usetikzlibrary{backgrounds}

\usepackage{pgfplots}
\pgfplotsset{compat=1.18}
\pgfplotsset{plot coordinates/math parser=false}

\tikzstyle{block} = [draw, fill=white, rectangle, 
    minimum height=1.6em, minimum width=3.2em]
\tikzstyle{sum} = [draw, circle, node distance=1cm]

\colorlet{lcfree}{Green3}
\colorlet{lcnorm}{Blue3}
\colorlet{lccong}{Red3}

\pgfdeclarelayer{marx}
\pgfsetlayers{main,marx}
\providecommand{\cmark}[2][]{\relax} 
\providecommand{\cmark}[2][]{%
  \begin{pgfonlayer}{marx}
    \node [nmark] at (c#2#1) {#2};
  \end{pgfonlayer}{marx}
  }

\providecommand{\keywords}[1]
{
  \small	
  \textbf{\textit{Keywords---}} #1
}

\usepackage[pdfa]{hyperref}
\hypersetup{%
    pdfauthor={Hasan Zakeri and Baisravan HomChaudhuri},
    pdftitle={Learning-Based Predictive Control Method for Vehicle Lateral Control with a Multi-Step Gaussian Process Regression Prediction},
}

\theoremstyle{definition}

\newcommand{\bx}{\ensuremath\boldsymbol x}
\newcommand{\bu}{\ensuremath\boldsymbol u}
\newcommand{\by}{\ensuremath\boldsymbol y}
\newcommand{\cX}{\ensuremath\mathcal{X}}
\newcommand{\cU}{\ensuremath\mathcal{U}}

\newcommand{\cD}{\ensuremath\mathcal D}

\makeatletter
\providecommand*{\diff}%
	{\@ifnextchar^{\DIfF}{\DIfF^{}}}
\def\DIfF^#1{%
	\mathop{\mathrm{\mathstrut d}}%
		\nolimits^{#1}\gobblespace}
\def\gobblespace{%
		\futurelet\diffarg\opspace}
\def\opspace{%
	\let\DiffSpace\!%
	\ifx\diffarg(%
		\let\DiffSpace\relax
	\else
		\ifx\diffarg[%
			\let\DiffSpace\relax
		\else
			\ifx\diffarg\{%
				\let\DiffSpace\relax
		\fi\fi\fi\DiffSpace}
\makeatother

\begin{document}
\author[1]{Hasan~Zakeri}
\author[2]{Baisravan~HomChaudhuri}
\affil[1]{
Mechanical, Materials, and Aerospace Engineering\\ Illinois Institute of Technology\\
Chicago, IL, USA\\
email: hzakeri@hawk.illinoistech.edu
}
\affil[2]{%
Member, ASME\\
Mechanical Engineering\\ Lamar University\\
email:~bhomchaudhur@lamar.edu 
}
\date{}
\title{Learning-Based Predictive Control Method for Vehicle Lateral Control with a Multi-Step Gaussian Process Regression Prediction}

\maketitle
\keywords{MPC, Gaussian Process Regression, Uncertainty Propagation, Vehicle Lateral Control, Lane Change}

\begin{abstract}
We introduce a novel approach to model predictive control that incorporates multi-step uncertainty prediction for safely controlling systems characterized by uncertainties dependent on both state and control variables. The discrepancy between real-world systems and their control-oriented representations arises from inherent uncertainties, which frequently correlate with state and control variables, a common occurrence in modeling errors. As these uncertainties accumulate and propagate over time, they can produce substantial deviations over extended horizons, potentially compromising the integrity of safety-critical applications. Although existing stochastic control frameworks can maintain system operation within safety boundaries at specified confidence levels, they necessitate accurate prediction of state distributions throughout the control horizon. This prediction represents a significant challenge for systems where uncertainties vary with state and control inputs. Our contribution addresses this challenge through a Model Predictive Controller leveraging multi-step Gaussian Process Regression to capture and anticipate uncertainties that are state- and control-dependent. We further propose an iterative solution to the optimization problem in our MPC framework and discuss the convergence of the algorithm. To demonstrate the method in a practical application, we conduct an in-depth analysis of vehicle lateral control, particularly during lane-changing maneuvers, examining how errors propagate through the system model. The effectiveness of our proposed methodology is validated through comprehensive simulations.
\end{abstract}

\section{Introduction}
The imperative for robust control systems in autonomous vehicles (AVs) stems from pressing transportation safety requirements. 
Contemporary AV frameworks utilize a hierarchical control structure comprising strategic path generation and tactical execution modules \cite{guanetti2018control,paden2016survey}. 
AV control strategies in transportation and navigation are often formulated as optimization problems, and Model Predictive Control (MPC) methods are well suited to handle them \cite{guanetti2018control, Mesbah.SMPC.2016}. 
MPC has been shown to handle complex stochastic systems with safety constraints, and is capable of real-time adaptation to dynamic environments~\cite{Mesbah.SMPC.2016}. 
Essential to these systems is probabilistic constraint satisfaction, particularly with regard to collision avoidance under operational uncertainties. The primary technical challenge lies in computing accurate state distribution forecasts, given the inherent complexities of nonlinear vehicle dynamics and stochastic perturbations in both state estimation and control implementation.

Vehicular systems inherently operate under diverse uncertainties, including parametric variations, state- and control-dependent disturbances, modeling approximations, numerical errors, and operational variations~\cite{limon2010robust}. 
These uncertainties manifest in multiple forms: system parameter ambiguity, stochastic noise in state and control variables, model simplification errors, numerical approximations, and dynamic utilization patterns. The collective impact of these uncertainty sources creates a notable discrepancy between the physical system and its control-oriented model, which we term ``mismatch'' or ``error.'' In predictive control frameworks, this mismatch exhibits temporal propagation characteristics, potentially resulting in substantial prediction errors over extended horizons. 
Such uncertainty propagation presents critical challenges for safety-guaranteed operations, necessitating careful quantification and integration into safety analysis. This work therefore focuses on developing a systematic approach to characterize error distributions across prediction horizons under state- and control-dependent uncertainties, subsequently incorporating these distributions into a novel MPC framework.

Gaussian Process Regression (GPR) \cite{hewing2018cautious, hewing2019cautious} represents a widely adopted methodology for modeling state- and control-dependent uncertainties. As a non-parametric approach, GPR characterizes functions through multivariate normal distributions, where function values at arbitrary input sets are jointly distributed with mean and covariance structures derived from training data. 
Existing methods primarily address instantaneous or single-step uncertainty quantification \cite{hewing2018cautious, hewing2019cautious}. 
This limitation proves particularly significant for stochastic predictive control applications, specifically stochastic model predictive control, where multi-step error distributions over finite horizons are essential. The capability to model only single-step uncertainties proves insufficient for these advanced control frameworks.

Even in linear systems with state- and control-dependent uncertainties, analytical derivation of future error distributions remains intractable~\cite{Mesbah.SMPC.2016}. 
This limitation has necessitated various approximation strategies in existing literature. 
Ning et al. \cite{ning_vehicle_2023} developed a learning-based extended kinematic model that incorporates states comparable to dynamic models, utilizing Gaussian Process (GP) models to quantify inter-model discrepancies. However, their approach only addresses immediate temporal discrepancies, neglecting the crucial aspect of uncertainty propagation relevant to model-based control frameworks. 
Alternatively, Broderick et al. \cite{broderick_dynamic_2012} proposed a comprehensive GP-based vehicle dynamics model integrated within an MPC framework. This approach, however, eschews valuable first-principle modeling insights in favor of black-box representations. The authors acknowledge that such models present challenges for safety-constrained control applications, necessitating additional approximations and simplifications to manage computational complexity.
Recent approaches have attempted to model error propagation through linearization of system dynamics and sequential propagation of single-step uncertainties \cite{hewing2018cautious, hewing2019cautious,hewing_cautious_2020}. 
However, this methodology introduces additional approximation errors in the propagation model. Alternative decision-making frameworks either rely on restrictive assumptions, such as linear dynamics with Gaussian uncertainty \cite{gao2014tube,carson2013robust}, or adopt overly conservative approaches by considering worst-case scenarios and over-approximating state distributions \cite{limon2010robust,paulson2018nonlinear, kousik2020bridging}. 
Such conservative approximations significantly constrain the solution space, leading to degraded AV performance in terms of goal achievement time and energy efficiency, and may even result in solution infeasibility \cite{chiang2017dynamic}. 
In this paper, we aim at developing a stochastic control strategy that focuses on reducing solution conservatism while ensuring probabilistic system constraint satisfaction. 

We focus on a lane change problem in this paper as an example case study.
Lane change maneuvers for AVs present significant challenges in motion planning and control due to coupled lateral and longitudinal nonlinear dynamics, non-convex collision avoidance constraints, and state- and control-dependent uncertainties. 
While numerous studies have addressed this problem \cite{bageshwar2004model,anderson2010optimal,liu2018dynamic,nilsson2016lane}, existing approaches exhibit notable limitations, often relying on pre-defined trajectories or very simple models for the vehicle. 
Recent advances in multi-vehicle coordination, including Buffered Input Cell (BIC) approaches \cite{zhou2017fast,wang2018safe,wang2019distributed}, primarily consider instantaneous vehicle positions rather than predicted trajectories, resulting in conservative constraints \cite{homchaudhuri2022distributed}. 
These challenges motivate the development of computationally tractable, non-conservative motion planning and control strategies that ensure safety under system uncertainties while enabling efficient lane changes.

This work addresses gaps in the existing literature by proposing a novel methodology to characterize and predict state- and control-dependent uncertainties across finite time horizons. The research introduces an innovative multistage GPR framework that extends beyond instantaneous uncertainty quantification to encompass temporal propagation dynamics. We have also proposed a novel stochastic MPC approach to guarantee vehicle safety based on the proposed uncertainty prediction model. 
Our primary contributions encompass three key elements: (1) the development of a multi-step GPR-based uncertainty model for systems with state- and control-dependent uncertainties, (2) the formulation of an MPC framework that leverages this probabilistic model to ensure robust safety guarantees, and (3) the introduction and implementation of a successive optimization approach and its convergence analysis to utilize multi-step GPR model to address non-convexities arising from state- and control-dependent uncertainties.
These theoretical advances are validated through comprehensive case studies focusing on autonomous vehicle lane change maneuvers.

In our previous work~\cite{zakeri_multi-step_2024}, we introduced a multi-step GPR strategy for learning uncertainty propagation. 
While that approach demonstrated promising results for basic lane-change scenarios, it lacked a theoretical foundation and extensive analysis.
The current paper extends this foundation by providing theoretical improvements and detailed analysis of our method that addresses these limitations. 
Specifically, we make the following new contributions: (i) a more comprehensive data gathering strategy, (ii) a general formulation for the MPC problem with detailed analysis including various comparisons, (iii) convergence analysis, and (iv) a scenario simulation.

The remainder of this paper is structured as follows: Section \ref{sec:problem_desc} presents the problem formulation, system dynamics, and introduces the dynamic model for our case study along with the safety-constrained MPC framework. Section \ref{sec:uncertaintyPropModel} details our methodology for uncertainty modeling and safety constraint formulation. The successive MPC approach and its convergence analysis are elaborated in Section \ref{sec:controller}. 
Numerical simulations are presented in Section~\ref{sec:numerical}, beginning with a benchmark inverted pendulum problem.
Section~\ref{sec:simulation} then provides comprehensive results for the lateral vehicle control case study, with emphasis on lane change scenarios. Finally, conclusions and future research directions are discussed in Section~\ref{sec:conclusion}.

\section{Problem Description}\label{sec:problem_desc}
We consider a system represented by
\begin{equation}\label{eq:actual_system}
\bx(k+1) = f(\bx(k), \theta, u(k), d(k)) + w(k)    
\end{equation}
where \(\bx(\cdot)\in \mathbb{R}^{n_x}\) is the state vector, \(u(\cdot)\in \mathbb{R}^{n_u}\) is the input to the system, and \(\theta\) 
{denotes the vector of physical system parameters (e.g., moment of inertia or mass), whose exact values might not be available. 
Finally, \(d\) and \(w\) are i.i.d. random process noises that are spatially uncorrelated. 
From now on, we use the term ``actual system'' to refer to the physical system (or an accurate physics-based model of the system).}
{We also employ a control-oriented model that can be used in the optimization and prediction stages, described by}
\begin{equation}\label{eq:controlModel}
x^\prime(k+1)=A(\hat{\theta})x^\prime(k)+Bu(k).
\end{equation}
{This model does not necessarily need to be linear, and any model, whether derived from first-principle modeling, input-output data, or model simplification, can be used here.}
The simple model may also depend on a set of parameters \(\hat{\theta}\), which is an estimate of \(\theta.\)
The mismatch between the actual system and the control-oriented model can be expressed as $ \varepsilon(k) = \bx(k)-\bx^\prime(k).$ 
{This modeling error is added at every prediction step and accumulates in the propagated uncertainty, resulting in significant error in long-horizon predictions.}
In this paper, {red}{we will take an in-depth look at modeling the uncertainty distribution over the entire prediction horizon \(H,\) where the uncertainty can depend on state and input, }
i.e., model $\rho(\tau), \tau\in[k,k+H]$ so that $\varepsilon (\tau) \sim \rho(\tau)$ where \(\varepsilon\) is defined as
\begin{equation}\label{eq:uncertainty}
\begin{aligned}
    \varepsilon(k+1) &= \bx(k+1)-\bx'(k+1)\\ 
    &= f(\bx(k),\theta,u(k),d(k))+w(k)-\left(A(\hat\theta)\bx'(k)+Bu(k)\right)
\end{aligned}
\end{equation}

It is worth noting that even though the system description Eq.~\eqref{eq:actual_system} does not show any dependence of the uncertainty on the system's states and inputs, the uncertainty as defined above will generally depend on state and input signals as shown in Eq.~\eqref{eq:uncertainty}.

\begin{figure}[bht]
    \centering
    \includegraphics[width=0.79\linewidth]{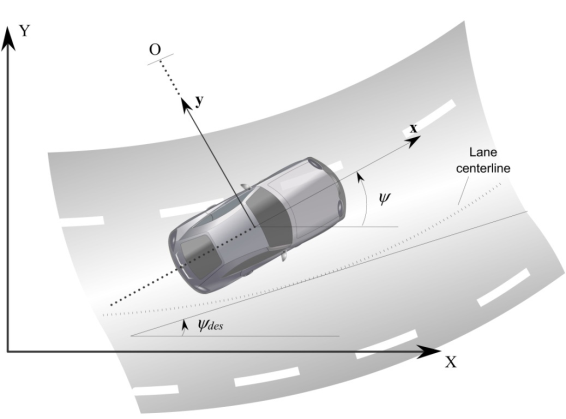}
    \caption{Lateral vehicle dynamics.}
    \label{fig:lateraldyn}
\end{figure}

\subsection{Lateral Vehicle Control System Dynamics}
Throughout this paper, we consider the lateral control of a vehicle as the case study, and use a bicycle model of the vehicle with two degrees of freedom,
as shown in Fig.~\ref{fig:lateraldyn}.
The vehicle's lateral position, \(y,\) is measured along the vehicle's lateral axis to point \(O,\) 
{which can be the center of rotation or another reference point/line.} 
The yaw angle, \(\Psi,\) is measured with respect to the global \(X\) axis. 
The longitudinal velocity of the vehicle at the center of gravity is denoted by \(V_x.\) 

The state-space model for the vehicle can now be written as \cite{rajamani_vehicle_2012}
{
\setlength{\arraycolsep}{3.5pt}
\begin{multline}
    \label{eq:dynamic_model}
    \frac{\diff}{\diff t}\begin{bmatrix}y\\\dot y\\\Psi\\\dot\Psi\end{bmatrix}=
    \begin{bmatrix}
        0&1&0&0\\0&-\frac{2C_{\alpha f}+2C_{\alpha r}}{mV_x}&0&-V_x-\frac{2C_{\alpha f}\ell_f-2C_{\alpha r}\ell_r}{mV_x}\\
        0&0&0&1\\
        0&-\frac{2\ell_fC_{\alpha f}-2\ell_rC_{\alpha r}}{I_zV_x}&0&-\frac{2\ell_f^2C_{\alpha f}+2\ell_r^2C_{\alpha r}}{I_zV_x}
    \end{bmatrix}
    \\\times\begin{bmatrix}y\\\dot y\\\Psi\\\dot\Psi\end{bmatrix}
    +\begin{bmatrix}
        0\\\frac{2C_{\alpha f}}{m}\\0\\\frac{2\ell_fC_{\alpha f}}{I_z}
    \end{bmatrix}\delta,
\end{multline}}
where \(C_{\alpha f}\) and \(C_{\alpha r}\) are the cornering stiffness of each front and rear tire, \(\ell_f\) and \(\ell_r\) are the distances of the front and rear tires from the center of gravity, $I_z$ is the moment of inertia about the \(z\) axis, and \(\delta\) is the front wheel steering angle.
In the above state-space model, the state vector is defined as $x=[y,\dot{y}, \Psi, \dot{\Psi}]^\top$
and the input is \(u(\cdot)=\delta(\cdot).\)
{See~\cite{rajamani_vehicle_2012} for details on the derivation of the model.}

The kinematic equations describing the lateral motion of the vehicle are expressed as
\begin{equation}\label{eq:kinematic}
\begin{aligned}
    \dot X &= V\cos(\Psi+\beta), &\dot Y &= V\sin(\Psi+\beta),\\
    \dot\Psi &= \frac{V}{\ell_f+\ell_r}\tan(\delta), &\dot V &= a
\end{aligned}
\end{equation}
where \(\Psi\) is the yaw angle (the orientation of the vehicle with respect to the global \(X\) axis), \(\beta\) is the vehicle sideslip angle, \(X\) and \(Y\) are the position coordinates, and \(\delta\) and \(a\) are the steering angle and acceleration (the two inputs to the system), respectively.
The velocity \(V\) is an external variable that can be obtained from a longitudinal model or assumed to be a time-varying function. 
The model in \eqref{eq:kinematic} will be utilized later to create potential trajectories for lane changes.

\subsection{Safety Constrained Model
Predictive Control}
We can formulate the lane change problem as a finite horizon optimal control problem. 
At every time step \(k,\) the optimization algorithm takes as inputs \(\bx(k), \bx_{ref}[k:k+H],\) and \(\bu_r[k:k+H],\) where \(\bx(k)\) is the current (measured) state of the vehicle, \(\bx_{ref}[k:k+H]\) is the desired and pre-defined reference trajectory, and \(\bu_r[k:k+H]\) is the input signal associated with the reference trajectory (reference input).
It then tries to solve the following optimization problem:
\begin{subequations}
\begin{align}
    \min&\sum_{t=k}^{k+H} \Vert\bx(t)-\bx_{ref}(t)\Vert_Q^2+(\bu_r(t)-\bu(t))^\intercal R(\bu_r(t)-\bu(t))\label{eq:optimizationproblem}
    \\
&\text{s.t. } \mathbb{P}(\bx(t)\in\mathcal{X}_{free}(t)) \geq 1-\epsilon_{safe} \label{eq:saferegion}
\end{align}
\end{subequations}
where $\Vert\bx(t)-\bx_{ref}(t)\Vert_Q^2 = (\bx(t)-\bx_{ref}(t))^\intercal Q(\bx(t)-\bx_{ref}(t))$, \(\cX_{free}(t)\) is the collision-free region at $t$, and $(1 - \epsilon_{safe})$ denotes our safety confidence.
This optimization problem keeps the vehicle as close to the desired trajectory as possible while ensuring that it stays in the collision-free region with a predefined degree of certainty $(1 - \epsilon_{safe})$.
It thereby solves the lane change problem: the reference trajectory supplies an ideal lane change path, and all obstacles (including other vehicles) and unsafe areas of the road are excluded from \(\cX_{free}(t).\)

Solving this problem with our multi-step uncertainty model is not straightforward: the future state distributions are outputs of multiple learning-based models, and the safety constraint of Eq.~\eqref{eq:saferegion} depends on the very input signal being optimized.
We return to these difficulties, and to our proposed solution, in Section~\ref{sec:controller}.

\section{Uncertainty Propagation Model} \label{sec:uncertaintyPropModel}
Modeling uncertainty, colloquially referred to as mismatch, recognizes the inherent discrepancies between our theoretical models and the complexities of real-world phenomena. 
It quantifies the deviation of the system’s actual behavior from its predicted behavior, thus allowing for a more robust and resilient optimization process. By accounting for this uncertainty, we enhance the method’s adaptability to unforeseen variations and disturbances. Most optimization methods rely on a simple model to ascertain safety; however, the mismatch between the model and the actual system can make the eventual trajectory unsafe. Our proposed method solves this issue by ensuring the simple model, augmented with the modeled uncertainty, stays in the safe region. 
This is done by shrinking the safety region by the amount of mismatch that we may have. 

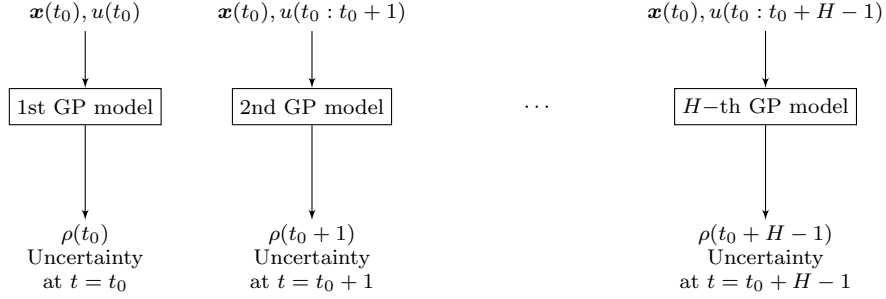
\begin{figure*}\small
    \centering
    \begin{tikzpicture}[auto, node distance=3cm,>=latex']
    \node [input, name=input1] {};
    \node [input, name=input2, right of=input1]{};
    \node [input,name=idots1, right of=input2,] {};
    \node [input, name=input3,right of=idots1] {};
    \node [block, below of=input1, node distance =1cm] (GP1) {\nth{1} GP model};
    \node [block, below of=input2, node distance =1cm] (GP2) {\nth{2} GP model};
    \node [block, below of=input3, node distance =1cm] (GP3) {$H-$th GP model};
    \node [name=dots1, below of=idots1,node distance =1cm] {$\cdots$};
    \node[output, below of=GP1,node distance =1.5cm] (out1){};
    \node[output, below of=GP2,node distance =1.5cm] (out2){};
    \node[output, below of=GP3,node distance =1.5cm] (out3){};
    \draw[draw,->] (input1)-- node[above,pos=0]{$\bx(t_0),u(t_0)$}(GP1);
    \draw[draw,->] (input2)-- node[above,pos=0]{$\bx(t_0),u(t_0:t_0+1)$}(GP2);
    \draw[draw,->] (input3)--node[above,pos=0]{$\bx(t_0),u(t_0:t_0+H-1)$}(GP3);
    \draw[draw,->] (GP1)-- (out1);
    \draw[draw,->] (GP2)-- (out2);
    \draw[draw,->] (GP3)-- (out3);
    \node[below of=out1,node distance =0.5cm,align=center] (unc1){%
    \(\rho(t_0)\)\\
    Uncertainty\\at $t=t_0$};
    \node[below of=out2,node distance =0.5cm,align=center] (unc2){%
    \(\rho(t_0+1)\)\\
    Uncertainty\\at $t=t_0+1$};
    \node[below of=out3,node distance =0.5cm,align=center] (unc3){%
    \(\rho(t_0+H-1)\)\\
    Uncertainty\\at $t=t_0+H-1$};
\end{tikzpicture}
    \caption{Data flow in the proposed method. A Gaussian Process Regression model is created for every time step of the horizon, representing the uncertainty at that time step.}
    \label{fig:ourwork_diagram}
\end{figure*}

{To overcome the challenge stemming from mismatch between} the actual system and the simple model for predictive controller synthesis, this paper focuses on developing a new multi-step Gaussian Process Regression (GPR) method, as shown in Fig. \ref{fig:ourwork_diagram}, {for predicting tight bounds on the uncertainty error in a prediction horizon.}
GPR is a non-parametric framework for nonlinear regression \cite{hewing2018cautious, hewing2019cautious}. 
A Gaussian process (GP) is a probability distribution over functions, such that every finite sample of function values is jointly Gaussian distributed.
Gaussian processes provide a principled, practical, probabilistic approach to learning in kernel machines, a broad class of supervised learning techniques.

Existing results in the literature focus on learning a one-step prediction error, \(\varepsilon(k+1)\), and train a GP to model this mismatch. 
The one-step error is then propagated over time through the system dynamics. 
This method can only approximate future mismatch, and assumes a linear and Gaussian propagation model. 
The novelty of our approach lies in modeling the mismatch, or prediction error, in every step, so we can have a higher confidence prediction with tighter bounds. 
Figure~\ref{fig:ourwork_diagram} shows how multiple GPRs are trained for each prediction horizon, which omits the approximate propagation steps and therefore results in a direct modeling of the uncertainty.
The method presented in this paper relies on the following steps:
\begin{enumerate}[label={\bfseries Step \arabic*}, leftmargin=4em, labelindent=1em, labelindent=!]
    \item Generate a set of inputs suitable for the system;
    \item apply the generated input signals to both the simple model and actual system, originating from the same initial condition;
    \item record the deviation between the two models as a function of initial condition and input signal, for every prediction step( from 1 to \(H\));
    \item train multiple GPR models, one for each time step over the horizon, on the mismatch error at every prediction horizon. 
\end{enumerate}
Here, we will explain every step in detail.

\subsection{Input Sequence Generation}
Accurate data collection begins with recognizing that the mismatch generally varies with both the state and the input. When only a single-step prediction of the error is required, drawing initial conditions and input signals at random from the admissible set is enough. The picture changes for longer horizons, as encountered in predictive control, where assembling the required data becomes unwieldy. The input signals must therefore be chosen deliberately, so that the input domain is not inflated beyond what is needed.

Covering every conceivable combination of inputs may seem prudent, yet only a comparatively small subset of them is ever meaningful. In the lane change study, for instance, a shift to the right lane corresponds to progressively steering right and then bringing the wheel back to center. Combinations that alternate erratically between left and right, or between high and low values, are not physically meaningful and are discarded; knowledge of the system is used instead to retain only the practical ones.

Realistic inputs are obtained here from the kinematic model of the system given in Eq.~\eqref{eq:kinematic}, whose inputs are the acceleration and the steering angle. Constructing a suitable input signal then exploits the differential flatness of that model, following~\cite{van1998differential,wang2018safe}.
Specifically, this kinematic model is differentially flat with respect to the flat output \(\sigma =\begin{bmatrix} \sigma_1 &\sigma_2 \end{bmatrix}^\intercal = \begin{bmatrix} x&y \end{bmatrix}^\intercal\), and the following output transformations can be used to determine a suitable set of inputs.
\begin{subequations}
\begin{align}
    x&=\sigma_1,&y&=\sigma_2\label{eq:flat1}\\
    \Psi&=\operatorname{atan2}(\dot\sigma_2, \dot\sigma_1), &v&=\sqrt{\dot\sigma_1^2+\dot\sigma_2^2}\label{eq:flat2}\\
    a&=\frac{\dot\sigma_1\ddot\sigma_1+\dot\sigma_2\ddot\sigma_2}{\sqrt{\dot\sigma_1^2+\dot\sigma_2^2}},&\delta&=\arctan\left(\frac{\dot\sigma_1\ddot\sigma_2-\dot\sigma_2\ddot\sigma_1}{(\dot\sigma_1^2+\dot\sigma_2^2)^{\tfrac{3}{2}}}(\ell_f+\ell_r)\right)\label{eq:flat_input}
\end{align}
\end{subequations}

A feasible and realistic trajectory for \(\sigma\), or \( \begin{bmatrix} x&y \end{bmatrix}^\intercal\), is built by taking \(y\) to be a fifth-order polynomial and \(x\) to be affine in time, with coefficients that follow from a set of conditions. Those conditions comprise the starting and target lanes, the speeds, and the desired heading of the car, which is required to align with the road. Enforcing them on the trajectory is what keeps it both feasible and realistic.
Considering the lateral position expressed as $y(t)=a_0t^5+a_1t^4+a_2t^3+a_3t^2+a_4t+a_5$,
the coefficients are obtained from the following conditions
\begin{subequations}
\begin{align}
y(t_0)&=y_0,& y(t_f)&=y_f,\label{eq:y0yf}\\
v_y(t_0)&=0,& v_y(t_f)&=0\label{eq:vy0vyf}\\
a_y(t_0)&=0,& a_y(t_f)&=0\label{eq:ay0ayf}
\end{align}
\end{subequations}
in which Eq.~\eqref{eq:y0yf} prescribes the lane change itself, with \(y_0\) the lateral position of the current lane and \(y_f\) that of the desired one. The pair in Eq.~\eqref{eq:vy0vyf} forces the lateral velocity to vanish at both ends of the maneuver, which is what keeps the car within its lane, while the last pair, Eq.~\eqref{eq:ay0ayf}, does the same for the lateral acceleration. The longitudinal speed is further taken to remain constant throughout the maneuver, so that \(x(t)\) is given by
\begin{equation}
x(t)=v_x(t-t_0)+x_0.
\end{equation}
A broad family of input signals then follows from Eq.~\eqref{eq:flat_input} by sweeping combinations of the initial and final lateral positions---or, more precisely, of the magnitude and direction of the lateral shift---together with the longitudinal speed and the final time.

\subsection{Data Generation and Collection}
Assume that a range of lane shifts, lane change times, and longitudinal speeds has been swept, yielding a collection of trajectories together with their feasible input sequences. 
Applying those sequences to the actual system, or to the simplified one, will not reproduce the planned trajectory exactly, since uncertainty intervenes. 
Their role is rather to supply a starting point from which plausible input signals can be generated.

The data generation starts by generating a set of possible inputs. We generate \(n_t\) different input signals.
For each input signal \(u^i(k)\in \mathbb{R}^{n_u}, i = 1,\dots, n_t, k=0,\dots,T_i-1\) generated, we obtain \(n_T\) different responses from the actual system by running this input signal \(n_T\) times, denoted as \(\bx_j^i(\cdot), i=1,\dots,n_t,j=1,\dots,n_T\), to observe the effects of randomness in the system model. 
Let the length of these responses also be \(T_i\). 
For each time instant over the horizon \(k=0,\dots,T_i-H,\) we use \(u^i(k:k+H)\) as the input signal and \(\bx_j^i(k)\) as the initial condition for the simple model to obtain the response \(\bx^{\prime i}_j(\cdot\mid v_x,\bx_j^i(k),u^i(k:k+H))\).

The error, or the mismatch, is given as
\begin{multline}
    \varepsilon_h(t\mid v_x,\bx_j^i(k),u^i(k:k+h))\\= \bx_j^i(t)-\bx^{\prime i}_j(t\mid v_x,\bx_j^i(k),u^i(k:k+h))
\end{multline}
for \(t=k+1,\dots,k+h\) and \(h=1,\dots, H.\) 
A sample data set of errors is thus available at each step of the prediction horizon.

For the lane change control problem the states are taken as \(x=\begin{bmatrix}y&\dot y&\Psi&\dot\Psi\end{bmatrix}^\intercal\) and \(x^\prime=\begin{bmatrix}y^\prime&\dot y'&\Psi'&\dot\Psi'\end{bmatrix}^\intercal.\) Starting from time step \(k,\) the error of the one-step prediction is then \(\varepsilon(k+1),\) that of the two-step prediction is \(\varepsilon(k+2),\) and in general the \(l\)-step prediction error \(\varepsilon(k+l)\) reads
\begin{equation}
\varepsilon(k+l)=\begin{bmatrix}
        y(k+l) - y'(k+l)\\
        \dot y(k+l)-\dot y'(k+l)\\
        \Psi(k+l)-\Psi'(k+l)\\
        \dot\Psi(k+l)-\dot\Psi'(k+l)
    \end{bmatrix}
\end{equation}
where \(\bx(k)=\bx'(k)\); that is, the simple model used for prediction is initialized at the current state of the actual system.

\subsection{Multi-Step Gaussian Process Regression}\label{sec:GP}

The goal in GPR is to infer the vector valued function \(\by=g(\bx)\) from previously collected (and possibly noisy) measurements of its input and outputs, or equivalently, a training set \(\cD\) of \(M\) observations, \(\cD=\{(\bx_i,\by_i),i=1,\cdots,M\}.\) Here, we will have \(H\) different GPR models for every prediction step, each using a radial basis function (RBF) kernel (also known as the squared-exponential kernel). See~\cite{rasmussen_gaussian_2006} for a more in-depth discussion on different choices of the kernel function. Having multiple GPR models for every prediction step allows us to have a more accurate representation of the uncertainties instead of propagating a one-step prediction through a model.

Once the error data is collected for various state and input sequences, the data is randomly split between a training and a validation set. 
For every prediction horizon \(h\), we train a GPR model with the input, at every sample, being \(\begin{bmatrix} v_x&\bx_j^i(k)^{\intercal} &u^i(k:k+h)\end{bmatrix}.\) 
The outcome is a family of \(H\) Gaussian process models, one per prediction step, denoted as \(G_h(\cdot).\) In practice, a separate model of this kind is trained for each state of the system. 
These Gaussian distributions have means and covariances that are functions of the state at current time instant and the input signal over the horizon.

\section{Successive Controller Design}\label{sec:controller}
\subsection{Formulation of the Optimization Problem}

{
}

A major challenge in ensuring safety is the dependence of uncertainty on the state and the input. Here, the input signal \(\bu(t)\) is our optimization variable. 
The constraints also depend on \(\bu(t).\) The constraints need to be modified at every step to account for the uncertainty as modeled by the multi-step GP model. 
However, this uncertainty also depends on \(\bu(t)\), which we do not know at the time of the optimization, and in general can be a nonlinear and non-convex function of \(\bu,\) making the whole optimization problem non-convex and intractable for real-time purposes.

Consider the optimization problem given as Eq.~\eqref{eq:optimizationproblem}. Adding system dynamics and input constraints to the problem, we obtain the following optimization problem:
\begin{subequations}\label{eq:opt}
\begin{align}
    \min\quad &\sum_{t=k}^{k+H} \Vert\bx(t)-\bx_{ref}(t)\Vert_Q^2
    + (\bu_r(t)-\bu(t))^\intercal R(\bu_r(t)-\bu(t))
    \label{eq:optcost}\\
    \text{s.t. }\quad&\bx(t+1)=A\bx(t)+B\bu(t)\\
    &\bu(t)\in\cU(t), \quad \bx(t)\in\mathcal{X}_{safe}(t)
    \label{eq:optsafety}\\
    &\mathcal{X}_{safe}(t) = \mathcal{X}_{free}(t) \ominus b(t)\\
    &b(t) = \{\bx\mid \Vert \bx-\mu_t\Vert\leq s\sigma_t\} \label{eq:const2}\\
    &GP_t(\bx(k),u(t))\sim\mathcal N(\mu_t,\sigma_t^2) \label{eq:shrinked}
\end{align}
\end{subequations}
where $\mathcal{X}_{free}(t)$ is the collision-free state set and $e_{model}(t)$ is the modeling error at $t$, distributed according to the GP of Eq.~\eqref{eq:shrinked}, i.e., $e_{model}(t)\sim\mathcal N(\mu_t,\sigma_t^2)$.
Here $b(t)$ is the back-off at time $t$, which is the $\epsilon_{safe}$ confidence interval of $e_{model}(t)$, \(s\) quantifies the span around the mean in the Gaussian distribution with the associated probability \(1-\epsilon_{safe}\),
and $\ominus$ is the Pontryagin difference.
The back-off $b(t)$ at the predicted time $t$ is computed based on the mean and covariance matrix of the uncertainty propagation model at $t$, so that the probabilistic safety constraint is satisfied.  
The constraint delineated by Eq.~\eqref{eq:const2} is designed to hold for a given confidence (with the probability of at least \(1-\epsilon_{safe}\)). 
Since our uncertainty is modeled to be Gaussian at every prediction step, we can translate this probability to a certain deviation from the mean in the Gaussian distribution.

Minimizing the cost in Eq.~\eqref{eq:optcost} with our multi-step uncertainty model is difficult for an optimizer because the future state distributions are outputs of multiple learning-based models.
Ensuring the safety of the vehicle through Eq.~\eqref{eq:optsafety} is also non-trivial, since this constraint depends on the prediction of future states, which in turn depends on \(\bu\).
The uncertainty itself is likewise a function of \(\bu\) and is therefore unknown before the optimization is solved, which renders the problem non-convex.
Additionally, our proposed uncertainty models are outputs of multiple GPR models and lack a simple closed form formula for an optimizer to use.
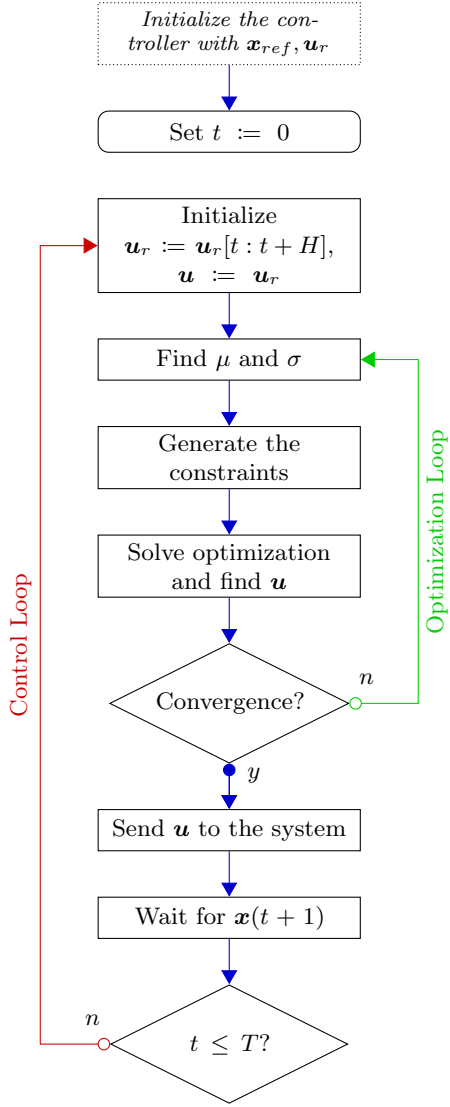
\begin{figure}[t]
    \centering
\begin{tikzpicture}[%
    >=triangle 60,
    start chain=going below,
    node distance=6mm and 60mm,
    every join/.style={norm},
    ]
\tikzset{
  base/.style={draw, on chain, on grid, align=center, minimum height=4ex},
  proc/.style={base, rectangle, text width=10em},
  test/.style={base, diamond, aspect=2, text width=6em},
  term/.style={proc, rounded corners},
  coord/.style={coordinate, on chain, on grid, node distance=6mm and 25mm},
  nmark/.style={draw, cyan, circle, font={\sffamily\bfseries}},
  norm/.style={->, draw, lcnorm},
  free/.style={->, draw, lcfree},
  cong/.style={->, draw, lccong},
  it/.style={font={\small\itshape}}
}
\node [proc, densely dotted, it] (p0) {Initialize the controller with \(\bx_{ref},\bu_r\)};
\node [term, join]      {Set \(t\coloneq0\)};
\node [proc] (p1) {Initialize\\ \(\bu_r\coloneq\bu_r[{t:t+H}],\)\\\(\bu\coloneq\bu_r\)};

\node [proc, join] (p2) {Find \(\mu\) and \(\sigma\)};
\node [proc, join] (p3) {Generate the constraints};
\node [proc, join] (p4) {Solve optimization\\and find \(\bu\)};
\node [test, join] (t1) {Convergence?};
\node [proc, join] (p5) {Send \(\bu\) to the system};
\node [proc, join] (p6) {Wait for \(\bx(t+1)\)};
\node [test, join] (t0) {\(t\leq T\/?\)};
\node [coord, right=of t0] (c0)  {}; \cmark{0}   
\node [coord, right=of t1] (c1)  {}; \cmark{1}   
\node [coord, right=of t1] (c1r) {}; \cmark[r]{1}
\node [coord, left=of t0] (c0l) {}; \cmark[l]{0}
\path (t1.south) to node [near start, xshift=1em] {$y$} (p5);
  \draw [*->,lcnorm] (t1.south) -- (p5);
\path (t1.east) to node [near start, yshift=1em] {$n$} (c1); 
  \draw [o->,lcfree] (t1.east) -- (c1) |- node[below,pos=0.25,rotate=90] {Optimization Loop} (p2);
\path (t0.west) to node [near start, yshift=1em] {$n$} (c0l); 
  \draw [o->,lccong] (t0.west) -- (c0l) |- node[above,pos=0.25,rotate=90] {Control Loop}  (p1);
\end{tikzpicture}
    \caption{Flowchart illustrating the successive approach to solving the optimization problem.}
    \label{fig:flowchart}
\end{figure}
The proposed successive controller synthesis approach focuses on successively solving the convexified optimal control problem, where the state distribution prediction over the horizon and the constraint tightening are done based on the control solution from the previous iteration. Since the constraints are now a convex function of the optimization variable, the new optimization program is convex.

The controller is first given a desired trajectory \(\bx_{ref},\) and a control trajectory \(\bu_r\) corresponding to it. 
This control trajectory comes from the path planner, which in this case relies on the differential flatness of the kinematic model Eq.~\eqref{eq:kinematic} and the associated transformation of Eqs.~\eqref{eq:flat1},~\eqref{eq:flat2}, and~\eqref{eq:flat_input}.
At each step of the control loop, the controller receives the car's current state, 
\(\bx(t)\) and starts a successive process at iteration \(j=1.\) This successive process is explained as follows. 
This state vector, along with the reference input 
\(\bu_r(t)\) serves as the initial condition for the GP models and helps approximate the modeling uncertainty, \(\rho(j,\bx(t),u^{j-1}_{[t:t+h]}),\)
which is a Gaussian distribution characterized by a mean 
\(\mu_h^j\) and a standard deviation \(\sigma_h^j.\) 
These statistical parameters are derived as \(\mu_h^j=\mu_h(\bx(t),\bu^{j-1}_{[t:t+h]})\) and \(\sigma_h^j=\sigma_h(\bx(t),\bu_{[t:t+h]}^{j-1}).\) Refer to Fig.~\ref{fig:flowchart} for a visualization of the algorithm.

These statistics are used to redefine and shrink the constraint according to Eq.~\eqref{eq:shrinked}. 
The optimization is then solved with this new set of constraints. 
At every iteration \(j,\) this new set of constraints is given by 
$\cX_{safe}^j(t)=\cX_{free}(t)\ominus b^{j-1}(t)$ where \(b^{j-1}(t)\) is the back-off with \(\epsilon_{safe}\) confidence interval of \(e_{model}^{j-1}[t:t+H]\) and \(e_{model}^{j-1}[t:t+H]\) is the output of the GP with the previous iteration's input, i.e., \(e_{model}\sim GP_t(x(t),u^{j-1}(t)).\)

However, since the shrinking of the constraints was based on an estimate of the input variables (or the previous iteration of the input), we use the results of this iteration as the starting point for finding the uncertainty and performing the optimization.
This successive approach continues until the convergence criterion is met.
The stopping criteria may be predicated on a diminution in the magnitude of changes observed in the objective function or the control variable \( u^j \). This ensures that the algorithm terminates when a satisfactory level of precision is attained, or when further iterations cease to yield significant improvements.

\subsection{Convergence Analysis}

The convergence of the successive solution approach, partially inspired by~\cite{maoSuccessiveConvexificationSuperlinearly2019}, hinges on two key conditions:
\begin{enumerate}
    \item An initial feasible control trajectory \(\bu_r\) exists such that \(\bu_r \in \mathcal{U}\) and the corresponding state trajectory \(\bx_r \in \mathcal{X}_{safe}^1\);
    \item At any iteration $j$, the previous solution remains feasible in the updated constraint set, i.e., $\bx^{*,j-1} \in \mathcal{X}_{safe}^j$ and $\bu^{*,j-1}\in \mathcal{U}$.
\end{enumerate}

The first condition is readily satisfied. The kinematic model~\eqref{eq:kinematic} with the differential flatness transformation in Eqs.~\eqref{eq:flat1}--\eqref{eq:flat_input} provides a preliminary control estimate \(\bu_r\) that, although not necessarily optimal, is based on the vehicle's equations of motion. The path planning algorithm ensures that this initial trajectory is feasible within the safety constraints, providing the starting point for the successive iterations.

For the second condition, a careful examination of the constraint set evolution is warranted. At iteration \(j,\) the safety set is \(\mathcal{X}_{safe}^j(t)=\mathcal{X}_{free}(t)\ominus b^{j-1}(t),\) where \(b^{j-1}(t)\) is computed from the GP model using \(\bu^{j-1}.\) Since the GP-based uncertainty model has bounded outputs, there exists a finite \(\varepsilon_G>0\) such that \(\Vert b^j(t)-b^{j-1}(t)\Vert\leq\varepsilon_G\) for all \(j\) and \(t.\) This means the change in the contracted constraint set between successive iterations is limited, and consequently \(\Vert \bu^{j}-\bu^{j-1}\Vert\leq u_{th}\) for some finite threshold \(u_{th}.\) With bounded changes in both the constraint set and the optimization variable, the state trajectory and control solution from the previous iteration remain feasible in the current iteration. See~\cite{mao2016successive,MAO20174063,maoSuccessiveConvexificationSuperlinearly2019} for a detailed treatment of feasibility preservation in successive convexification.

Given these two conditions, the convergence argument is as follows. At iteration \(j,\) since the previous solution \((\bx^{*,j-1}, \bu^{*,j-1})\) is feasible under the updated constraints, it serves as a feasible candidate for the current optimization. Because the optimizer minimizes the cost over this feasible set, the new optimal cost can only be less than or equal to the previous one:
\[
    J^j(\bx(k), \bx_{ref}, \bu_r) \leq J^{j-1}(\bx(k), \bx_{ref}, \bu_r).
\]
That is, each iteration either yields an improved solution or retains the previous one. Additionally, since the cost function is a sum of quadratic terms with positive semidefinite weight matrices \(Q\) and \(R,\) it is non-negative, providing a lower bound. A non-increasing sequence that is lower-bounded must converge, establishing that the successive approach converges toward a local minimum.

\section{Numerical 
Simulations}\label{sec:numerical}
To demonstrate the effectiveness of our proposed approach for uncertainty quantification, we present two comprehensive numerical simulation studies followed by a comparison against the state-of-the-art. 
The first study is the inverted pendulum system, which has simpler dynamic equations, enabling a deeper dive into the intricacies of the method. The second is a realistic vehicle control problem.

The inverted pendulum provides an ideal testbed due to its nonlinear dynamics and sensitivity to model uncertainties, allowing for systematic evaluation of how GP regression captures and propagates parametric uncertainties and unmodeled dynamics. The well-understood behavior of this system enables a clear assessment of the GP framework's ability to learn and represent model discrepancies from limited data, making it an appropriate benchmark to validate the proposed uncertainty quantification approach.

\subsection{Benchmark Problem: Inverted Pendulum}\label{subsec:pendulum}
The inverted pendulum system is simplified to consist only of a rigid massless rod of length \(l\) with a point mass at its tip, pivoting about a fixed point. Assuming no friction or any resistance to movement, the dynamics is governed by the nonlinear equation
\begin{equation}\label{eq:pend:nonlinear}
    \ddot\theta = \frac{g}{l}\sin\theta, 
\end{equation}
where \(\theta\) represents the angular displacement from the upright position. For control design purposes, the system is often linearized around the unstable equilibrium \(\theta = 0,\) which yields
\begin{equation}\label{eq:pend:linear}
    \ddot\theta = \frac{g}{l}\theta.
\end{equation}
It is assumed that our knowledge of the length has \(\pm10\%\) error, and our measurements have an additive noise of \(\mathcal N(0, 0.02).\) To generate training data for the GP uncertainty model, simulations are performed from multiple initial conditions using both the full nonlinear and linearized models under identical control inputs, with the resulting trajectory differences serving as observations of the model discrepancy that the Gaussian process will learn to approximate. 

The simulation study is conducted over an 8-step time horizon (with a sampling time of \(T=\qty{0.1}{\second}\)), comparing two distinct GP-based uncertainty modeling approaches. In the proposed multi-step GP method, a separate Gaussian process is trained at each time step using the accumulated model discrepancies observed up to that point, directly capturing the time-varying nature of the uncertainty propagation. This is contrasted with a conventional one-step GP approach that trains a single model on immediate step-ahead prediction errors and relies on iterative propagation of uncertainty estimates through successive time steps (called the ``propagation method'' for short). The comparative analysis evaluates how effectively each method captures and represents the evolution of model uncertainty over the prediction horizon.

The collected data is split between a training set and a test set by randomly choosing a portion of the initial conditions for the test set, and allocating any trajectory generated from them as the test set to avoid contamination of the training set. 
Because the pendulum length perturbation is drawn afresh for each trajectory and is not an input to the regression, the discrepancy at a given initial condition is genuinely random rather than a deterministic function of the inputs. 
Estimating that spread therefore requires {repeated} trajectories from each initial condition: the training set uses \(20\) independent trajectories per initial condition, which allows the noise term of the kernel to be identified separately from the underlying function. 
Training with a single trajectory per initial condition leaves the noise term to be inferred from scatter across different initial conditions, which biases it low and produces predicted intervals that are too narrow.
For each test condition, Monte Carlo simulations are performed using both the multi-step and one-step GP approaches to generate probabilistic predictions of the model discrepancy over the 8-step horizon. 
The resulting uncertainty estimates from both methods are then compared against the actual observed differences between the actual system and linearized models on the test trajectories, evaluating each approach's ability to accurately quantify and propagate uncertainty in out-of-sample scenarios.
To show the effectiveness of any uncertainty prediction method, three metrics are considered at every step of the prediction horizon: coverage, mean accuracy, and over-approximation.

The first metric, \emph{coverage}, asks whether the predicted distribution actually contains the realized error. It is measured by counting the percentage of Monte Carlo trajectories that fall inside the prediction margin. 
The prediction margin here is selected as \(2\sigma,\) so that approximately \(95\%\) of all trajectories are expected to fall in this confidence region. 
Coverage substantially below the nominal level means 
under-approximation,
and any constraint tightening based on it would be unsafe.
Coverage alone is not sufficient, because a model that over-approximates the true uncertainty can achieve near-perfect coverage while 
leading to conservative results and even infeasibility. 
The remaining two metrics therefore quantify how 
accurate and tight the prediction is. 
Both are normalized by the same quantity, the Monte Carlo sample standard deviation \(s_i,\) so that they can be read together.

The second metric, \emph{mean accuracy}, compares the predicted mean to the sample mean of the trajectories. Since there are multiple initial conditions in the test set and each can have a large or small standard deviation, a normalized mean squared error (with a formulation similar to that of the reduced \(\chi^2\) statistic) is used:
\begin{equation}\label{eq:nmse}
    \text{NMSE} = \frac{1}{N}\sum_{i=1}^{N}\left(\frac{\hat{\mu}_i - \bar{\mu}_i}{s_i}\right)^2
\end{equation}
where $N$ is the number of test initial conditions, $\hat{\mu}_i$ is the predicted mean for initial condition $i$, $\bar{\mu}_i$ and and $s_i$ are the sample mean standard deviation from Monte Carlo simulations, respectively. 
An \(\text{NMSE}\) of \(a\) means the predicted mean is \(\sqrt{a}\) sample standard deviations away from the observed mean.

The third metric is the \emph{over-approximation ratio} (OAR), defined as the
average ratio of the predicted standard deviation to the Monte Carlo
sample standard deviation:
\begin{equation}\label{eq:oar}
    \text{OAR} = \frac{1}{N}\sum_{i=1}^{N}\frac{\sigma_i}{s_i}
\end{equation}
An ideal uncertainty model would yield $\text{OAR} = 1$, indicating that
the predicted uncertainty matches the observed spread exactly.
Values significantly greater than $1$ indicate over-approximation,
which leads to conservative constraints and potential infeasibility
in the MPC framework. Values below $1$ indicate under-approximation, which is unsafe.

\begin{figure}
    \centering
    \includegraphics[width=\linewidth]{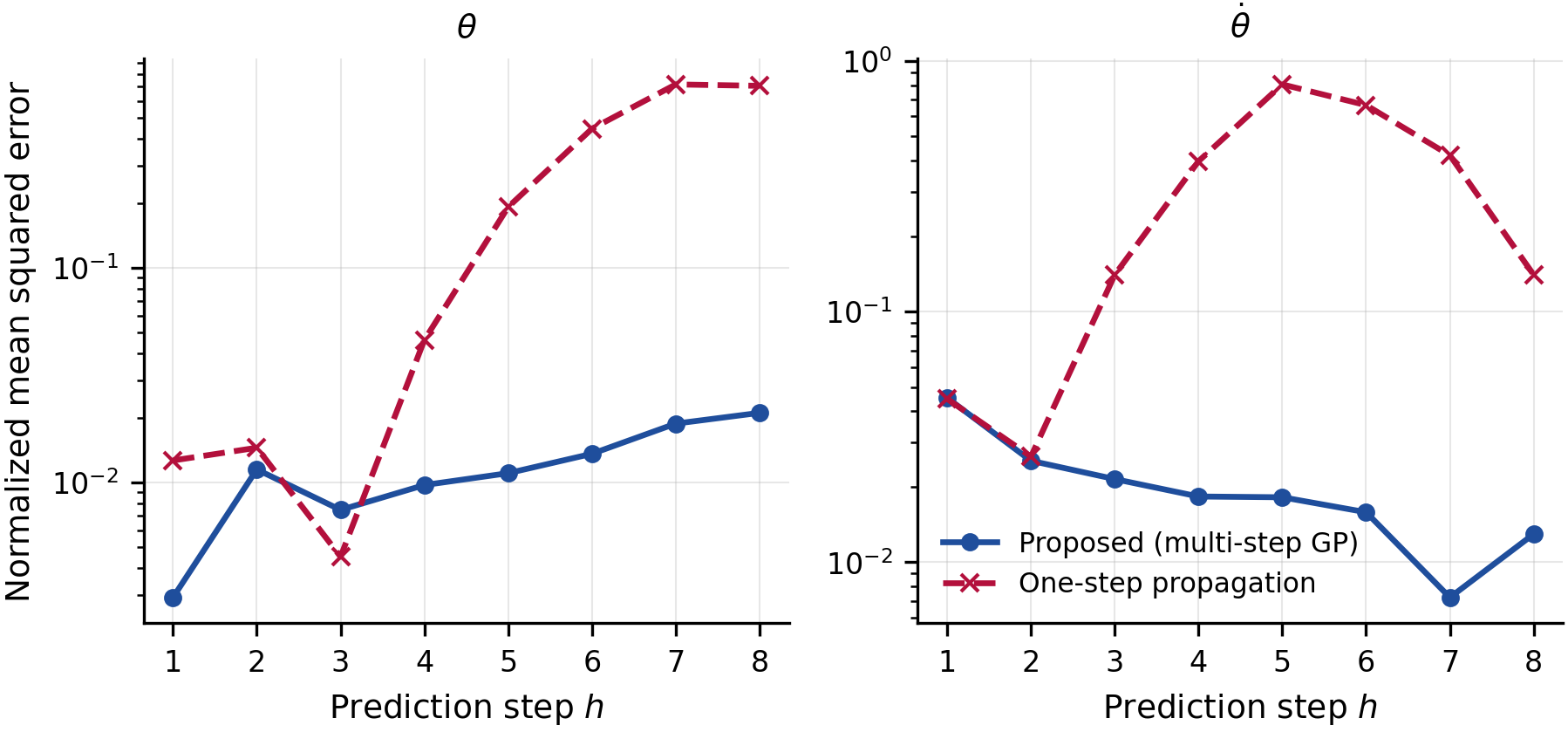}
    \caption{Normalized mean squared error, per Eq.~\eqref{eq:nmse}, comparing the proposed multi-step GP method and the propagation method over the 8-step prediction horizon, for the angular displacement \(\theta\) (left) and the angular rate \(\dot\theta\) (right). Note the logarithmic vertical scale.}
    \label{fig:pendulum_mse}
\end{figure}
\begin{figure}
    \centering
    \includegraphics[width=\linewidth]{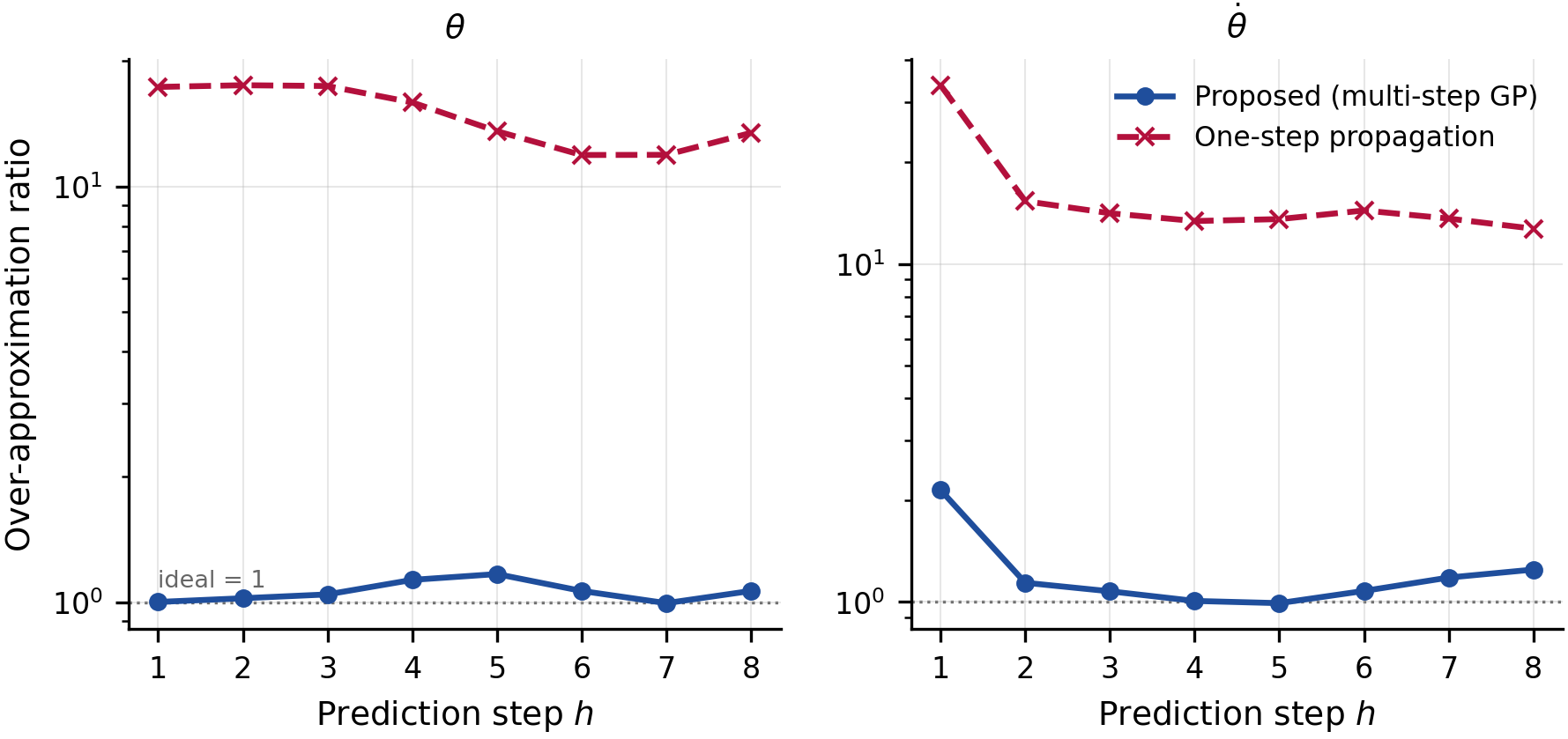}
    \caption{Over-approximation ratio, per Eq.~\eqref{eq:oar}, for \(\theta\) (left) and \(\dot\theta\) (right). The dotted line marks the ideal value of unity. The proposed method tracks the observed spread almost exactly, while the propagation method over-approximates it by more than an order of magnitude. Note the logarithmic vertical scale.}
    \label{fig:pendulum_overapprox}
\end{figure}

The results for the inverted pendulum benchmark are presented in Figs.~\ref{fig:pendulum_mse} and~\ref{fig:pendulum_overapprox}, in both cases for the angular displacement \(\theta\) and the angular rate \(\dot\theta\).
Figure~\ref{fig:pendulum_mse} shows the NMSE over the 8-step prediction horizon. The two methods agree closely at the first two or three steps, where a single propagation step is still a good approximation. Beyond that the propagation method's error grows by more than an order of magnitude, reaching \(0.72\) for \(\theta\) at step 7 and \(0.81\) for \(\dot\theta\) at step 5, whereas the proposed method remains between \(0.01\) and \(0.02\) across the entire horizon for both states. Averaged over the horizon, the proposed method's NMSE is \(0.01\) against \(0.27\) for \(\theta\), and \(0.02\) against \(0.33\) for \(\dot\theta\). This demonstrates that the proposed method avoids the accumulation of propagation errors in predicting the mean of the uncertainty distribution.

Figure~\ref{fig:pendulum_overapprox} shows the over-approximation ratio for both methods. The propagation method consistently over-approximates the uncertainty by more than an order of magnitude, averaging \(14.9\times\) for \(\theta\) and \(16.4\times\) for \(\dot\theta\) across the horizon. The proposed method stays close to the ideal value of unity, averaging \(1.06\times\) and \(1.23\times\) respectively---an improvement of \(14.0\times\) and \(13.3\times\) over the propagation method.
Crucially, this tightness is not obtained at the expense of safety. Over the same test set the proposed method retains \(95.9\%\) coverage of the nominal \(95\%\) interval for both states, so the predicted tube contains the realized error as often as it is designed to while being an order of magnitude smaller. It is also worth noting that the propagation method does not benefit from the additional training data: its over-approximation is a structural consequence of compounding a one-step covariance through the horizon, not a matter of estimation quality.
The proposed method's lower over-approximation translates directly to less conservative control actions---a critical advantage for safety-constrained MPC, where excessively tight constraints can lead to infeasibility.

\subsection{Lateral Control and Lane Change Scenario}
\label{sec:simulation}
\begin{figure}%
    \centering%
    \includegraphics[width=0.75\linewidth]{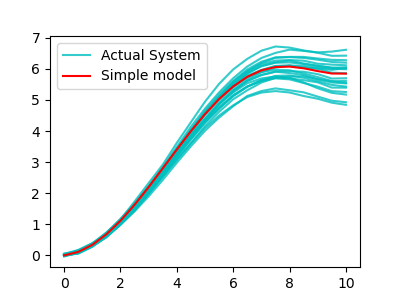}%
    \caption{Lateral position \(y\) from the simple model, together with 20 responses from the actual system for the same input sequence and initial condition. The horizontal axis is the time step \(k\) (with \(\Delta t=\qty{0.5}{\second}\)).}
    \label{fig:sample_response}
\end{figure}

To validate our proposed approach, we consider a lane-change scenario where the actual system is governed by Eq.~\eqref{eq:dynamic_model}. 
Here, the parameters are given as (refer to~\cite{haddad_density-based_2022,bakker_new_1989}) $\ell_f=\qty{1.432}{\meter},\ell_r=\qty{1.472}{\meter},        m=\qty{2050}{\kilogram}, I_z=\qty[inter-unit-product = \ensuremath{{}\cdot{}}]{3344}{\kilogram\meter\squared}, 
        C_{\alpha f}=\qty[inter-unit-product = \ensuremath{{}\cdot{}}]{250}{\kilo\newton\per\radian}, C_{\alpha r} =\qty[inter-unit-product = \ensuremath{{}\cdot{}}]{250}{\kilo\newton\per\radian}$.
The safety region of Eq.~\eqref{eq:saferegion} is defined as 
\begin{equation}
    \Vert\bx'(t)-\bx_r(t)\Vert \leq r(t),\quad\forall k<t\leq k+H
\end{equation}
where \(\bx'\) denotes the set of state values admitted by the uncertainty.
The control-oriented model in \eqref{eq:dynamic_model} is considered to have parameter error of 5\%, and at every step, the states are further corrupted by Gaussian measurement noise. 
\(V_x\) is also considered to be uncertain and is randomly perturbed. 
Fig.~\ref{fig:sample_response} illustrates how these perturbations affect the response of the simple model, alongside 20 trajectories produced by the actual model from identical inputs and initial conditions.

Training data for the multi-step GP models is produced from
lateral shifts of \(\Delta y = 3.2\) and \(\Delta y = 6.4\) in both directions, longitudinal speeds drawn from \(\{15, 20, 25, 30\}\), and final times \(t_f\) of \(6\), \(7\), and \(8\) seconds.
This data is then split into training and test sets, and the multi-step GP models are fit to the error.
Note that although this example starts from a linear model, nothing in the method depends on linearity, on other structural assumptions about the system, or on any simplification, which is what makes it well suited to predictive control.
The quantitative comparison against the propagation method of~\cite{hewing_cautious_2020} is carried by the benchmark study of Section~\ref{subsec:pendulum}; for the vehicle model we demonstrate the practical consequence directly. 

\begin{figure}[tbh]
    \centering
    \includegraphics[width=0.8\linewidth]{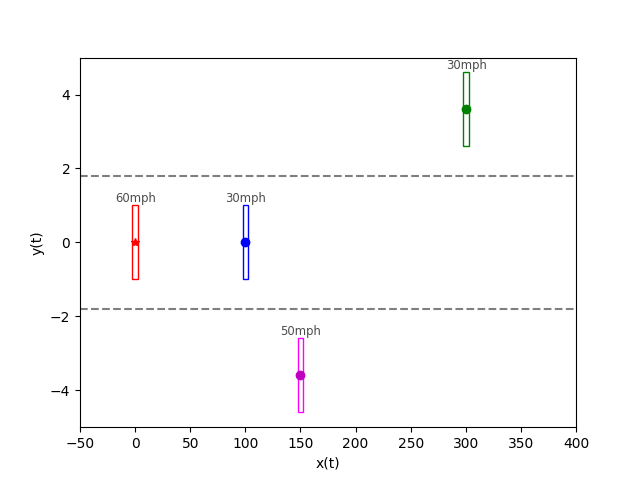}
    \caption{The beginning of the scenario. The longitudinal speed of each vehicle is specified in miles per hour (mph). The vehicle under control, or the ego vehicle, is depicted in red (left most car in the middle lane). The dot marks the center of gravity of each vehicle and the surrounding rectangle shows its physical size; the dashed lines are the lane boundaries.}
    \label{fig:scenario}
\end{figure}

\begin{figure}[tbh]
    \centering
    \includegraphics[width=0.8\linewidth]{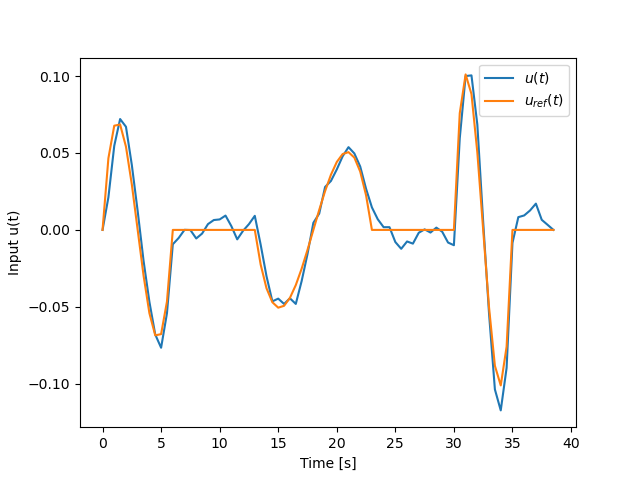}
    \caption{Steering angle input from the MPC controller ($u(t)$, blue) compared to the reference input ($u_{ref}(t)$, orange) over the lane change scenario.}
    \label{fig:scenario.U}
\end{figure}

\begin{figure*}%
    \centering%
    \includegraphics[width=0.7\linewidth]{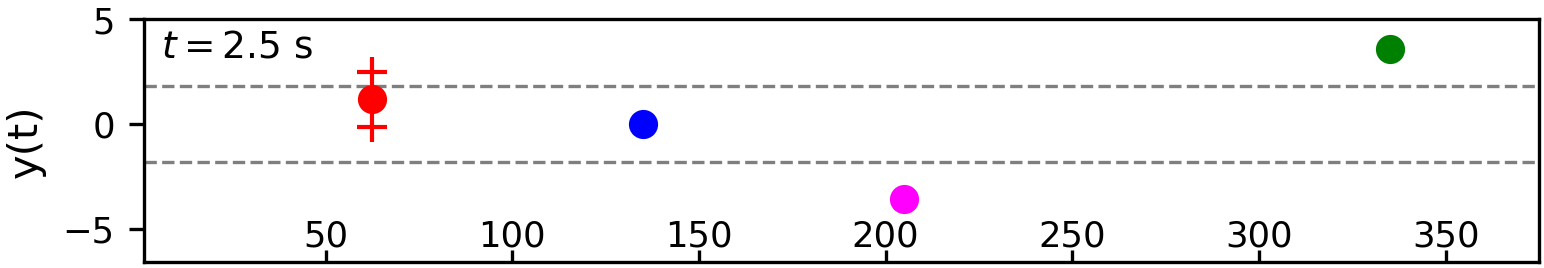}\\
    \includegraphics[width=0.7\linewidth]{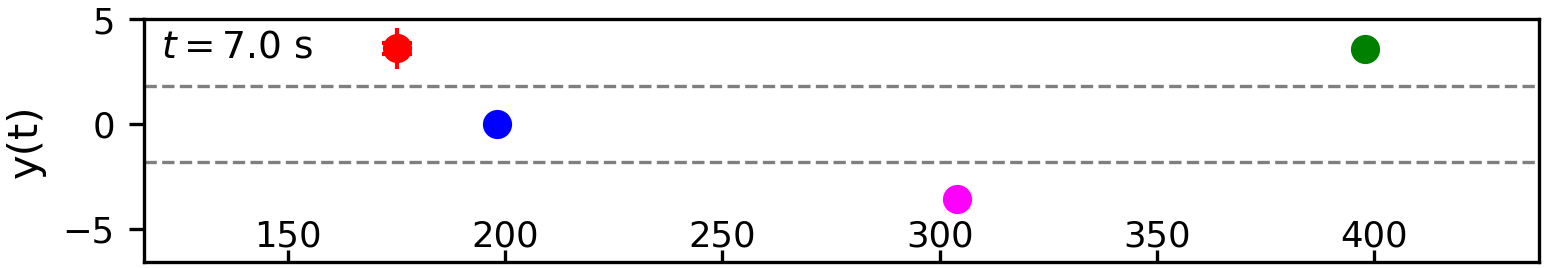}\\
    \includegraphics[width=0.7\linewidth]{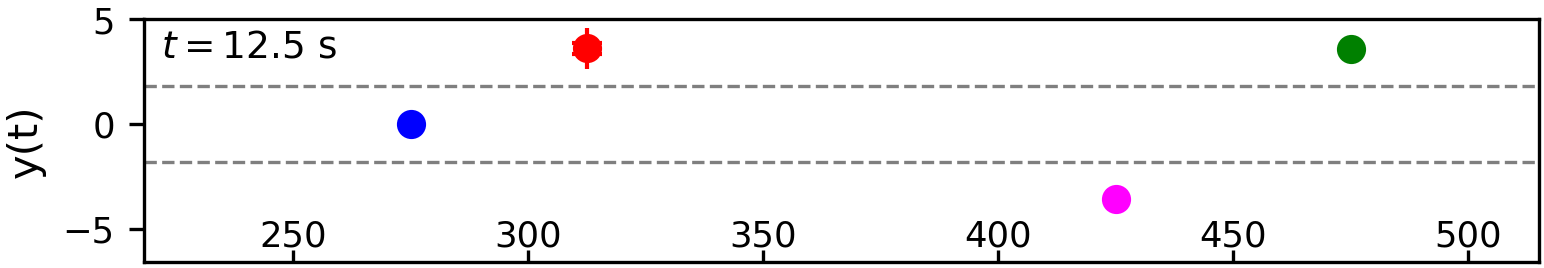}\\
    \includegraphics[width=0.7\linewidth]{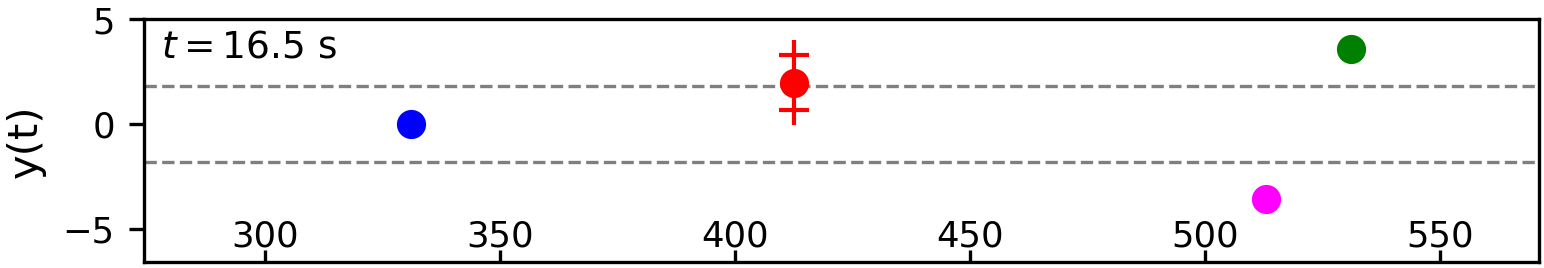}\\
    \includegraphics[width=0.7\linewidth]{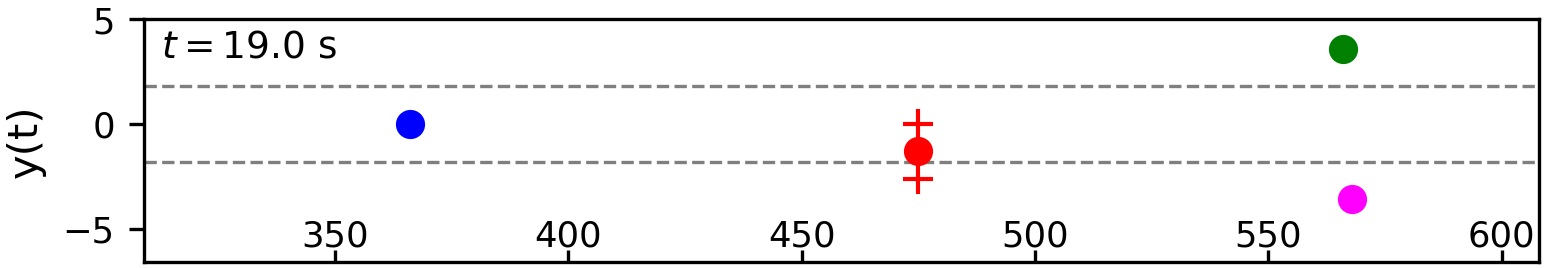}\\
    \includegraphics[width=0.7\linewidth]{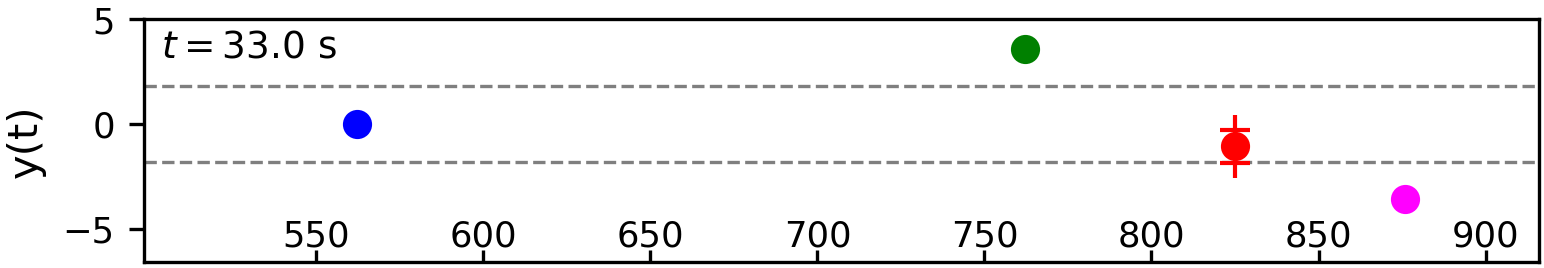}\\
    \includegraphics[width=0.7\linewidth]{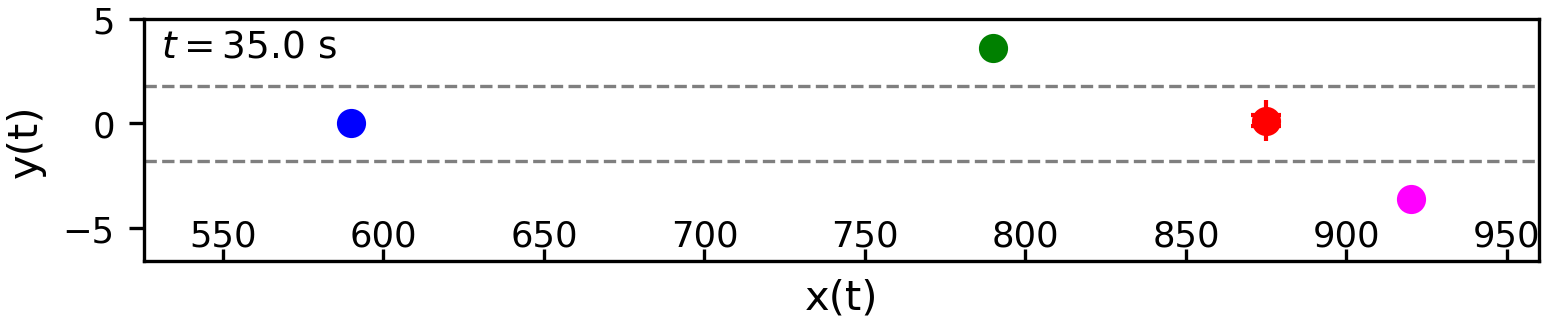}
\caption{Position of the ego vehicle relative to the other vehicles on the road at seven instants, with the elapsed time marked in each panel. The dots depict the center of gravity of each vehicle, the ego vehicle is shown in red, and the dashed lines are the lane boundaries. The plus marks above and below the ego vehicle delimit its collision-free zone, which is defined with respect to the center of gravity of the vehicle rather than its physical bounds.}
\label{fig:scenario.time}
\end{figure*}

Having established the accuracy of the multi-step uncertainty prediction in Section ~\ref{subsec:pendulum}, we apply the successive controller of Section~\ref{sec:controller} to a full driving scenario.
The start of a common lane-change scenario is depicted in Fig.~\ref{fig:scenario}, where the longitudinal speed of each vehicle is specified in miles per hour (mph).
The vehicle must make a series of lane changes to maintain a safe distance from surrounding vehicles that are moving more slowly.
The output of the proposed control solution is shown in Fig.~\ref{fig:scenario.U}, while
Fig.~\ref{fig:scenario.time} shows the position of vehicles at different times during the simulation.
The ego vehicle first leaves the middle lane for the upper lane (\(t=\qty{2.5}{\second}\) and \(t=\qty{7}{\second}\)) and holds it while passing the slower traffic (\(t=\qty{12.5}{\second}\)). It then crosses back down toward the lower lane (\(t=\qty{16.5}{\second}\) and \(t=\qty{19}{\second}\)) and finally settles back into the middle lane as the road ahead clears (\(t=\qty{33}{\second}\) and \(t=\qty{35}{\second}\)).
We can see that our algorithm handles the lane changes effectively while maintaining safety.

\begin{figure}[h!]
    \centering%
    \includegraphics[width=0.9\linewidth]{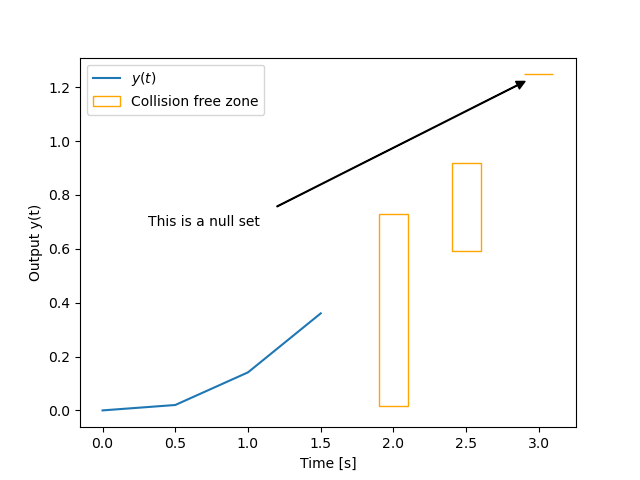}
    \caption{The moment at which uncertainty propagation through one-step estimation methods fails. The blue line depicts the closed-loop response so far (i.e., the trajectory the car has taken), which ends at \(t=\qty{1.5}{\second}\). The orange boxes are the collision-free zones for the three predicted steps; the third collapses to a null set at \(t=\qty{3}{\second}\), showing how overestimation of the uncertainty leads to an infeasible problem. 
    }
    \label{fig:scenario_fail2}
\end{figure}

\begin{figure}[h!]
    \centering
    \includegraphics[width=\linewidth]{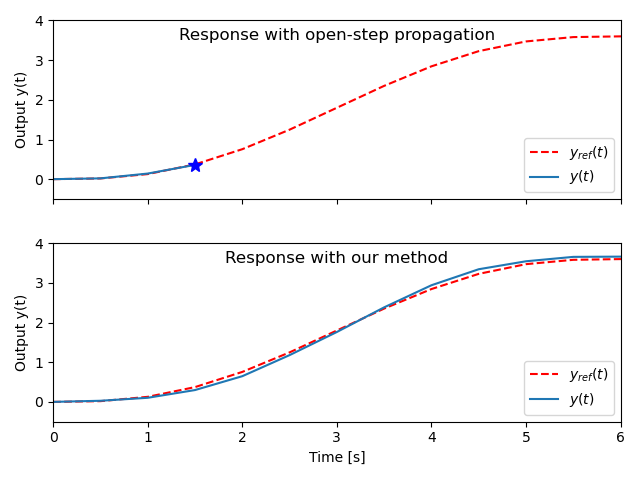}
    \caption{(Top) The state-of-the-art one-step propagation method fails to follow the trajectory after the first few steps due to over-constrained approximation; the star marks the instant at which the problem becomes infeasible. Our method (bottom) manages to navigate the same trajectory with the same parameters successfully.}
    \label{fig:scenario_fail3}
\end{figure}

We compared the impact of our approach with the uncertainty modeling and propagation method proposed in \cite{hewing_cautious_2020} in the lane-change problem.  
To make an effective comparison, both methods have been trained on the same data (for the one-step prediction). 
While in several cases the uncertainty modeling approach in \cite{hewing_cautious_2020} is adequate for controller development, it lead to null feasible control constraints in tight safe regions.  
Due to over-approximation, the method in~\cite{hewing_cautious_2020} fails to 
find a solution if the collision-free zone is small (as shown in Fig. ~\ref{fig:scenario_fail3}), since its feasible set becomes a null set (as shown in Fig. ~\ref{fig:scenario_fail2}). 
As discussed earlier, the constraint set is constructed by calculating the set difference between the safety zone and the approximated uncertainty.
Figure~\ref{fig:scenario_fail2} shows the moment at which the feasible set for controller leads to a null constraint set due to the conservative nature of the one-step propagation approach. 
The controller found a solution at every step up to \(t=\qty{1.5}{\second}\); solving at that instant over a three-step horizon, however, the constraint set predicted for \(t=\qty{3}{\second}\) is empty, and no feasible input exists.
The orange boxes in Fig. \ref{fig:scenario_fail2} are the collision-free zones for the three predicted steps, obtained by shrinking the intended safety zone by the predicted uncertainty. At the third step the uncertainty margin of the propagation method of~\cite{hewing_cautious_2020} exceeds the safety zone entirely, collapsing the zone to the null set marked in the figure.
Figure~\ref{fig:scenario_fail3} shows that the state-of-the-art method fails to follow the trajectory after the first few steps due to this over-constrained approximation. 
On the other hand, our proposed method, since it aims to reduce over-approximation of the uncertainty propagation, didn't face this issue and resulted in the vehicle maneuver safely under the tighter constraints.

All simulations are performed on a personal laptop with an Intel 
Core i7 CPU and 16 GB of RAM. The controller's computational burden is low, allowing real-time execution. The training of the GP requires more computational overhead, but as it is done offline and in advance, this does not pose a challenge to the applicability of the method. The training is done in Python 3.10. 

\section{Conclusion}\label{sec:conclusion}

{The paper introduces a novel approach for uncertainty modeling in dynamic systems through multi-step Gaussian Process Regression for state- and control-dependent uncertainties. Unlike existing methods that rely on single-step estimations, our framework addresses the complex challenge of tracking uncertainty propagation across all prediction horizons. We have also discussed the convergence of the algorithm. 
The vehicle lateral control case study demonstrated the practical utility of our approach during lane change maneuvers. The simulation results revealed significant advantages over traditional uncertainty approximation techniques, particularly in preventing the excessive conservatism that typically arises from over-approximation in extended prediction horizons. The improved accuracy in uncertainty prediction directly translates into enhanced safety assurances while minimizing performance compromises. This balance of safety and performance optimization represents a meaningful contribution to safety-critical control systems, with potential applications extending beyond automotive systems.}

\section*{ACKNOWLEDGMENT}

This material is based on work supported by the National Science Foundation (grant $\#2130718$).

\bibliographystyle{ieeetr}
\bibliography{Docs/refs.bib} 
\balance
\end{document}